\documentclass{aa}  
\usepackage{graphicx}
\usepackage{longtable}
\usepackage{multirow}
\usepackage{lscape}
\usepackage{pdflscape}
\usepackage{caption}
\usepackage{rotating}
\usepackage{natbib}
\usepackage{amsmath}
\usepackage{mathtools}
\usepackage{float}
\usepackage[above,below]{placeins}

\usepackage{txfonts}
\newcommand{\kms}{km\,s$^{-1}$}       %km/s

\newcommand{\cmthree}{cm$^{-3}$}
\newcommand{\gcmtwo}{g cm$^{-2}$}
\newcommand{\um}{$\mu$m}                                 %micron
\newcommand{\msunyr}{M$_{\odot}$\,yr$^{-1}$}

\newcommand{\gl}{\;\lower.6ex\hbox{$<$}\kern-7.75pt\raise.65ex\hbox {$>$}\;}
\newcommand{\lapprox}{$\stackrel {<}{_{\sim}}$}
\newcommand{\higal}{Hi-GAL}
\newcommand{\amin}{$^{\prime}$}                   %arcus and coordinates

\newcommand{\adeg}{$^{\circ}$}

\newcommand{\hii}{H{\sc ii}}

\begin{document}
\title{Large-scale latitude distortions of the inner Milky Way Disk from the Herschel/Hi-GAL Survey.}

%   \subtitle{}

\author{S. Molinari \inst{1} \and
        A. Noriega-Crespo \inst{2} \and 
        J. Bally \inst{3} \and
        T. J. T. Moore \inst{4} \and
        D. Elia \inst{1} \and
        E. Schisano \inst{1} \and
        R. Plume \inst{5} \and
        B. Swinyard \inst{6} \and
        A. M. Di Giorgio \inst{1} \and
        S. Pezzuto \inst{1} \and
        M. Benedettini \inst{1} \and
        L. Testi \inst{7, 8} 
        }

\institute{INAF- Istituto di Astrofisica e Planetologia Spaziale, Via Fosso del Cavaliere 100, I-00133 Roma, Italy
       \email{molinari@iaps.inaf.it} \and
	Space Telescope Science Institute, 3700 San Martin Dr., Baltimore, 21218 MD, USA \and
	Center for Astrophysics and Space Astronomy, University of Colorado, Boulder, 80309 CO, USA \and
	Astrophysics Research Institute, Liverpool John Moores University, Liverpool Science Park Ic2, 146 Brownlow Hill, Liverpool L3 5RF, UK \and
	Department of Physics \& Astronomy, University of Calgary, Canada \and
	STFC, Rutherford Appleton Labs, Didcot, UK \and
	European Southern Observatory, Karl Schwarzschild str. 2, 85748 Garching, Germany \and
	INAF-Osservatorio Astrofisico di Arcetri, Largo E. Fermi 5, 50125 Firenze, Italy 
	        }

%          \fnmsep\thanks{}

%   \offprints{}

   \date{Received ; accepted}

  \abstract
   {}
   {We use the {\textit Herschel} \higal\ survey data to study the spatial distribution in Galactic longitude and latitude of the interstellar medium and of dense, star-forming clumps in the inner Galaxy. }
   {We assemble a complete mosaic of the inner Galaxy between $l = -$70\adeg\ and +68\adeg\ in the far-infrared continuum from \higal. The peak position and width of the latitude distribution of the dust column density is analysed by fitting a polynomial function to the diffuse IR surface brightness in 1\adeg\ longitude bins and the result is compared to MIPSGAL 24-\um\ data. The latitude distribution of number density of compact sources from the band-merged \higal\ photometric catalogues is also analysed as a function of longitude.}
   {The width of the diffuse dust column density traced by the \higal\ 500-\um\ emission varies across the inner Galaxy, with a mean value of 1\adeg \!.2$-$1\adeg \!.3, similar to that of the distribution of MIPSGAL 24-\um\ sources and of \higal\ sources with a 250-\um\ counterpart. \higal\ sources with a 70-\um\ counterpart define a much thinner disk, with a mean FWHM$\sim$0\adeg\!.75, in excess of the result obtained by the ATLASGAL submillimetre survey. The discrepancy with the 250-\um\ source distribution can be explained by relatively higher confusion in the {\textit Herschel} data in the midplane region. The peak of the average latitude distribution of \higal\ sources is at $b\sim -$0\adeg \!.06, coincident with the results from ATLASGAL. The detailed latitude distribution as a function of longitude shows clear modulations, both for the diffuse emission and for the compact sources. The displacements are mostly toward negative latitudes, with excursions of $\sim$0\adeg\!.2 below the midplane at $l \sim$ +40\adeg, +12\adeg, $-$25\adeg\ and $-$40\adeg. The only positive bend peaks at $l\sim -$5\adeg. No such modulations can be found in the MIPSGAL 24\um\ or WISE 22\um\ data when the entire source samples are considered; modulations in part similar to the ones exhibited by the Herschel sources appear when the mid-infrared catalogues are filtered according to criteria that preferentially select YSOs.}
   {The distortions of the Galactic inner disk revealed by {\textit Herschel} confirm previous findings from CO surveys and HII/OB source counts but with much greater statistical significance and are interpreted as large-scale bending modes of the Plane. The lack of similar distortions in tracers of more evolved YSOs or stars rules out gravitational instabilities or satellite-induced perturbations, as they should act on both the diffuse and stellar disk components. We propose that the observed bends are caused by incoming flows of extra-planar gas from the Galactic fountain or the Galactic halo interacting with the gaseous disk. Stars, having a much lower cross-section, decouple from the gaseous ISM and relax into the stellar disk potential. The timescale required for the disappearance of the distortions from the diffuse ISM to the relatively evolved YSO stages are compatible with star-formation timescales. }
   

	\authorrunning{Molinari et al.}
	\titlerunning{Latitude distortions in the Milky Way Disk from \higal }
   \maketitle
%
%________________________________________________________________

\section{Introduction}
\label{intro}

% general intro - the existence of corrugations is well known:

The detection of low-amplitude "corrugations" in the disk of the Milky Way, within the radius where the full HI warp starts to develop, dates back at least to \cite{Gum+1960}, who found that the inner Galactic disk, while very flat, contains localised excursions of around 20\,pc from the principal Plane of the Galaxy.  \cite{Quiroga1974} reported large-scale latitude modulations in the distribution of the HI emission and OB associations as a function of longitude. This result was confirmed by \cite{Lockman1977} and, specifically for the Milky Way Central Molecular Zone, by \cite{LB1980}.  
Such corrugations have also been reported for the disks of other spiral galaxies, e.g., by \cite{MU2008} for IC 2233, with a more pronounced amplitude for \hii\ regions and the star-forming component in general, and basically undetectable in the older stellar component traced by mid-IR continuum radiation. More recently, \cite{McClure+2012} used higher-quality radio data to confirm the scenario proposed by \cite{LB1980} of an HI distribution in the CMZ organised in a tilted elliptical disk with an inclination $\sim$24\adeg .

%% potential generating mechanisms:

It has been suggested that macroscopic disk distortions, such as the HI warps commonly seen in spiral galaxies as well as in the Milky Way, may be caused by the gravitational action of orbiting dwarf satellites, like the Magellanic Clouds in the local system. Close passages or minor mergers of the dominant galaxy with such minor systems may well have produced warps (e.g., Sagittarius A in the case of the Milky Way). Although HI warps generally occur in the external regions of spiral galaxies, beyond the radius of the stellar disk, N-body modelling by Edelsohn and Elmegreen (1997) predicted that a satellite the size of the LMC can also generate height and perpendicular velocity perturbations in the inner disk.  On the other hand, \cite{Weinberg1991} concluded that only long-wavelength modes such as the Galactic warp would be excited by such interactions.  In addition, 
\cite{Pranav+Jog2010} showed that the Galactic potential in the inner regions of the disk is intense enough to suppress the development of perturbations induced by dwarf satellites, while the perturbation would be free to fully develop into the observed large-scale warps only at larger radii where the stellar density drops. 

%% importance - relation to SF directly and indirectly via refuelling:

The observed structures may, however, be more than just interesting details of the mechanics of spiral galaxies and may in fact be related to the star-formation process.
Alfaro et al (1992) have suggested that corrugations seen in the Sagittarius-Carina arm are related to spatially correlated enhancements in star-formation activity, with a causal connection via 3D waves and the growth of gravitational and magnetic instabilities.  
\cite{Franco+1988} investigated the possibility that impacts on the Galactic Plane by High-velocity Clouds (HVCs) may explain the high latitude of the Orion and Monoceros star-forming regions.  The effect of magnetic fields was added by \cite{Santi+1999}, the conclusion of which was that the HVC gas does not penetrate the inner disk but can induce oscillations and trigger Parker instabilities.  Such interactions may result in significantly enhanced gas densities in previously magnetically subcritical clouds \citep{Vaidya+2013}.  The Parker instability itself may give rise to undulations in spiral arms in both the azimuthal and vertical directions, and in the $z$-component of the velocity, while also producing very large gas concentrations with masses similar to those of HI super clouds in the Galaxy \citep{Franco+2002}.

%  No clear correlation between SFR and oscillation maxima in the Hi-Gal data?

%Hopkins et al (2012?) predict that accretion flows onto the disk will increase the scale height of the disc gas (but not the turbulent velocity dispersion) so expect scale height to be enhanced where the distortions are greatest?  There may be a correlation between FWHM and median offset in the Hi-Gal data at positive longitudes but not obviously so at negative ones.

What is generally termed ``Galactic fountain" gas, is Galactic ISM material pushed out of the Plane with sufficient momentum by supernova explosions in OB associations, that then falls back onto the Plane after cooling \citep{Spitzer1990}, perhaps in the form of HVCs, as suggested by \cite{Bregman1980} and \cite{Kwak+2009}.  

Accretion onto galaxies from halo gas, either from gas stripped from orbiting dwarf galaxies or accretion from a diffuse halo, is commonly invoked to solve the gas depletion problem in star-forming galaxies (e.g., \citealt{Sancisi+2008}). \cite{Peek2009} simulates gas accretion from different sources in the Galactic potential and concludes that accretion from a diffuse halo is able to channel fresh ISM onto the most currently active star-forming Galactocentric radii, while accretion of gas stripped from dwarf satellites would be dominant outside the Solar circle, therefore requiring efficient inward gas radial transport. Although the latter is possible in principle, the need to transfer angular momentum outward would seem to make the process very inefficient, even in the presence of well developed spiral arms (e.g., \citealt{Peek2009}).  In addition, \cite{Marinacci+2010} model the effect of the passage of fountain gas through hot coronal gas, and find that this can cause the latter to condense and cool locally to provide an adequate gas supply rate.

% also mention IVCs seen in Planck?

While the effects of these phenomena are clearly fundamental to the maintenance of star formation in the Galaxy, the direct detection of this inflowing gas is still elusive.  The high-latitude clouds revealed in HI \citep{Wakker+1991} and in the infrared continuum and CO line emission \citep{Blitz+1984, Weiland+1986} may be participating in this, but it is not at all clear to what extent they provide a reliable estimate of the entire budget of diffuse outer gas currently accreting onto the Milky Way. 
One possibility would be to reveal these accreting gas flows indirectly through their dynamical effects on the Galactic disk. Given the cross sections involved, the effects should be more detectable on the diffuse component of the disk. The latest generation Galactic Plane surveys in the infrared and submillimetre provide ideal datasets to analyse the large-scale morphological properties of the disk in its diffuse and cold ISM phase.

The numerical magnitude of the Hi-GAL source catalogues with respect to previous submillimetre surveys like the Bolocam Galactic Plane Survey (BGPS \citealp{Rosolowsky2010}) or ATLASGAL \citep{Contreras2013, Csengeri+2014}, allows us to analyse the latitude distribution of sources with unprecedented detail and statistical significance as a function of longitude. Furthermore, the sensitivity of the {\it Herschel} PACS and SPIRE cameras coupled with optimal map-making image reconstruction developed for \higal\ \citep{trafi11,Piazzo+2012} allows us to also characterise in great detail the latitude distribution of the diffuse emission from the Galactic ISM. The analysis below is based on the first release of \higal\ maps and photometric catalogues of compact sources in which the large majority has sizes from point-like to twice the instrumental beam \citep{Molinari+2015b}. Additional processing done here is the production of multi-tile mosaics for the SPIRE 500-\um\ band to allow precise quantitative measurement of the latitude distribution of the ISM thermal dust emission.  This step was carried out using the new Unimap \citep{Piazzo+2015} software and represents an upgrade of the \higal\ pipeline that will be used for further data releases.

\section{The panoramic view of the inner Galaxy.}
\label{distr_glat}

%The numerical magnitude of the Hi-GAL source catalogues with respect to previous submillimeter surveys like BGPS \citep{Rosolowsky2010} or ATLASGAL \citep{Contreras2013, Csengeri+2014} allows us to analyse the latitude distribution of sources with unprecedented detail and statistical significance as a function of longitude. Furthermore, the sensitivity of the {\it Herschel} PACS and SPIRE cameras coupled with optimal map-making image reconstruction developed for \higal\ \citep{trafi11,Piazzo+2012} allows us to also characterise in great detail the latitude distribution of the diffuse emission from the Galactic ISM. The analysis below is based on the first release of \higal\ maps and source catalogues \citep{Molinari+2015b}. Additional processing done here is the production of multi-tile mosaics for the SPIRE 500-\um\ band to allow precise quantitative measurement of the latitude distribution of the ISM thermal dust emission.  This was produced using the new Unimap \citep{Piazzo+2014} software that represents the upgrade of the \higal\ pipeline and that will be used for further data releases.

\begin{sidewaysfigure}[t]
\vspace{0 cm}
\includegraphics{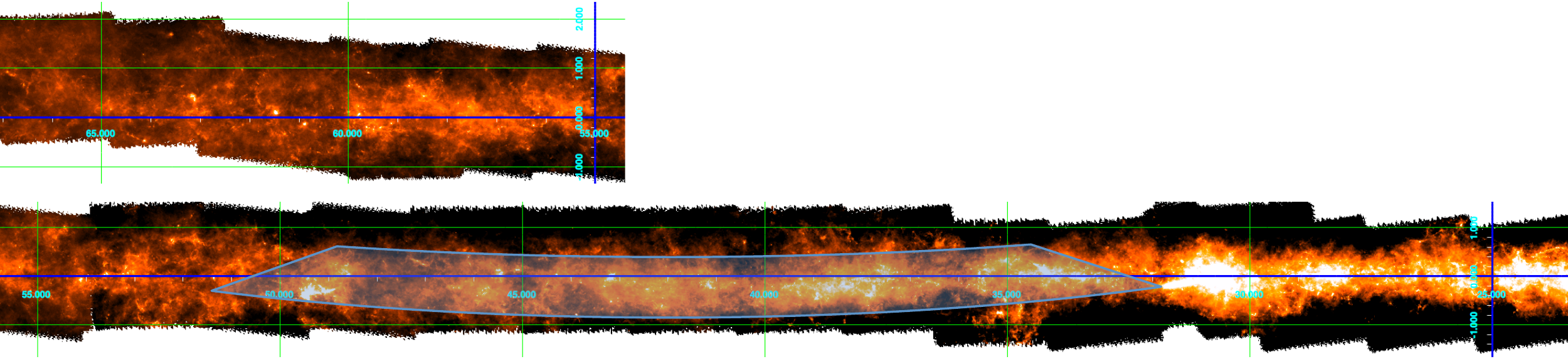}
\caption{\textbf{\textit{a)}} 500-\um\ panoramic mosaic of the Galactic Plane for the longitude range 54\adeg $\leq\ l \leq$ 67\adeg\ (upper panel)   and 23\adeg\!.5 $\leq\ l \leq$ 56\adeg\ (lower panel). The longitude and latitude scales are given in the figures. The horizontal blue line marks $b$=0\adeg, while the green-line grid marks intervals of 5\adeg\ in longitude and 1\adeg\ in latitude. The lightly shaded areas are intended to visually emphasise the slow latitude modulations of the overall dust thermal emission.}
\label{barpan}
\end{sidewaysfigure}

\addtocounter{figure}{-1}
\begin{sidewaysfigure}[t]
\vspace{0 cm}
\includegraphics{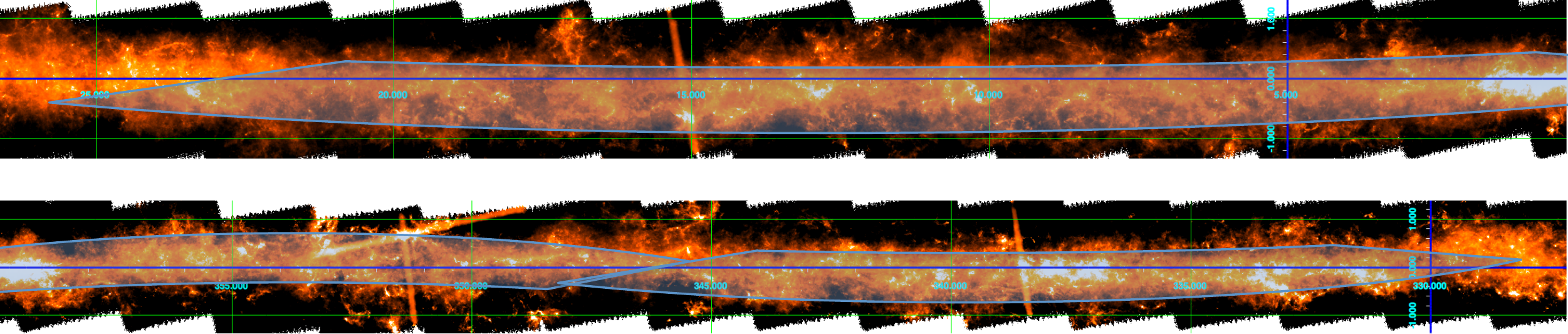}
\caption{\textbf{\textit{b)}} As fig. \ref{barpan}a, but for the longitude ranges 0\adeg $\leq\ l \leq $ 27\adeg\ (upper panel) and 327\adeg $\leq$ 360\adeg\ (lower panel), encompassing the range spanned by the Galactic Bar.}
%\label{q1pan}
\end{sidewaysfigure}

\addtocounter{figure}{-1}
\begin{sidewaysfigure}[t]
\vspace{0 cm}
\includegraphics{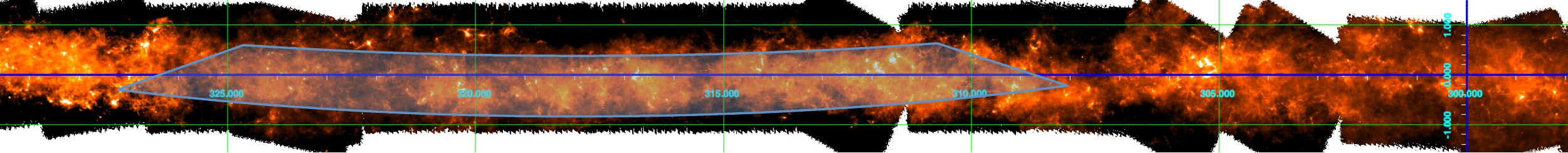} 
\caption{\textbf{\textit{c)}} As fig. \ref{barpan}c, but for the longitude range 298\adeg $\leq\ l \leq$ 329\adeg .}
%\label{q4pan}
\end{sidewaysfigure}

\FloatBarrier

\subsection{The latitude distribution of the diffuse ISM.}
\label{sect_distr_ism}

\begin{figure}
\centering
\includegraphics[width=0.5\textwidth]{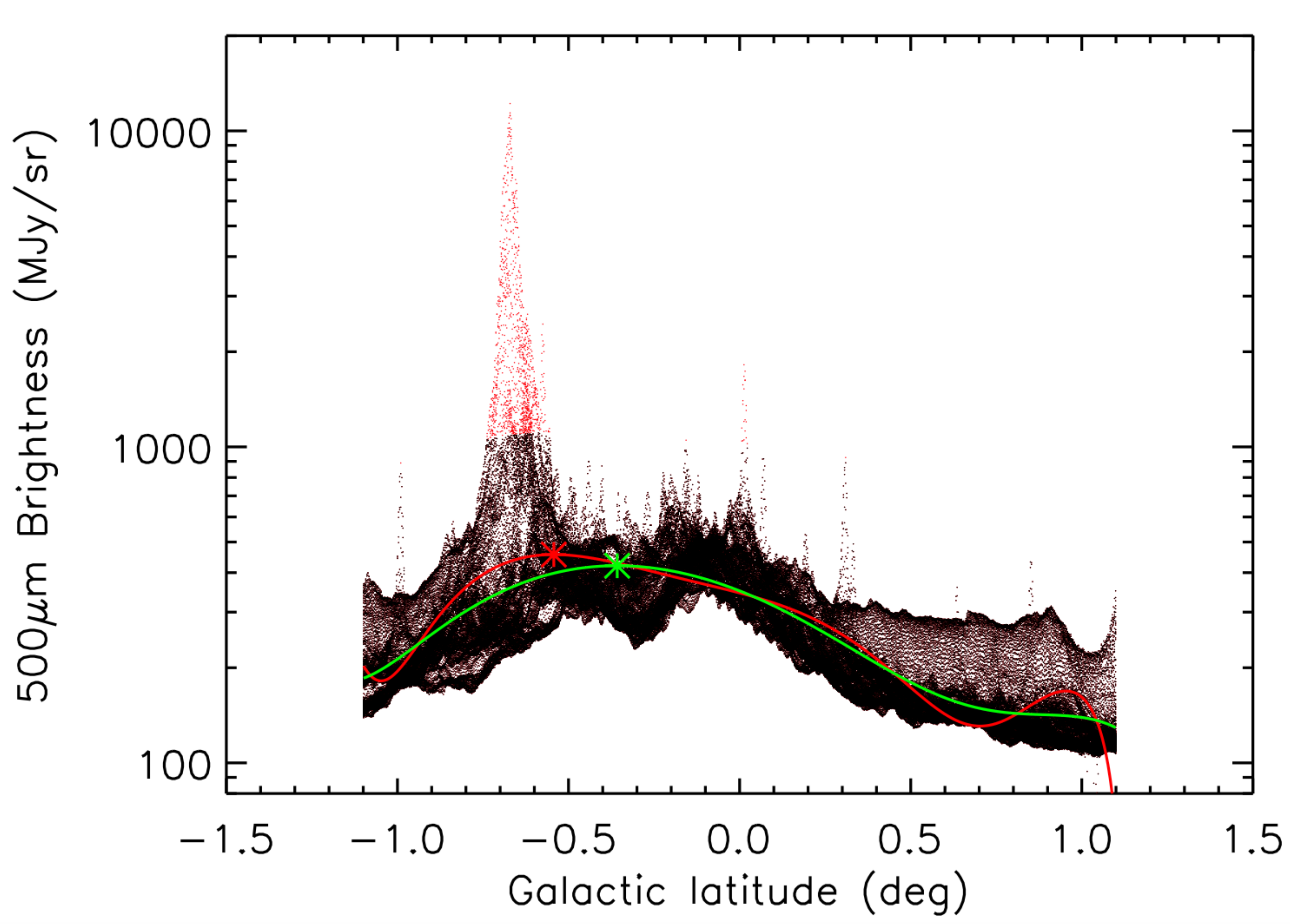}
\caption{500-\um\ brightness values (all the points) as a function of Galactic latitude in the longitude interval 14\adeg\!.5 $\leq l \leq$ 15\adeg\!.5 from the map in fig. \ref{barpan}b (this longitude range contains M16, which is clearly visible as the large bump at $b\sim$ $-$0\adeg\!.7). The red thick line is the 7$^{th}$-order polynomial fit to all points, and the red points are those that are clipped away as they lie above 3 times the r.m.s. of the residuals (after subtracting the fit). The green thick line is the 5$^{th}$-order polynomial fit to all points that survived this clipping (all the black points). The thick crosses mark the positions of the maxima of each respective fit.}
\label{cliptest}
\end{figure}

In fig. \ref{barpan}a, b and c, we present panoramic mosaics of the \higal\ 500-\um\ emission, together covering most of the inner Galaxy surveyed by {\it Herschel} and subject of the present initial data release. 
%The mosaics have been obtained with the \textit{Unimap} map-making package \cite{Piazzo+2014} using the entire 500-\um\ data set. 
These panoramic views are extremely useful to get a qualitative impression of the latitude distribution of the dust thermal emission and identify systematic patterns. These 500-\um\ panoramic mosaics can also be used in a quantitatively more rigorous way by estimating the latitude of the centroid emission as a function of longitude. Since bright source complexes (large star-forming regions) that could be offset with respect to the overall emission distribution (see, e.g., M16/M17 in fig. \ref{barpan}b would bias the estimate of an overall Galactic Plane centroid of emission, the first step is to clip away these complexes. We first subdivide the image in bins of 1\adeg\ amplitude in longitude and, using a 7$^{th}$-order polynomial, fit the brightness values of all pixels in each longitude bin as a function of latitude (thick red line in fig. \ref{cliptest}, where we illustrate the particular case of the 1\adeg\ bin centered at $l=15$\adeg\ that contains M16). 

\begin{figure}
\centering
\includegraphics[width=0.5\textwidth]{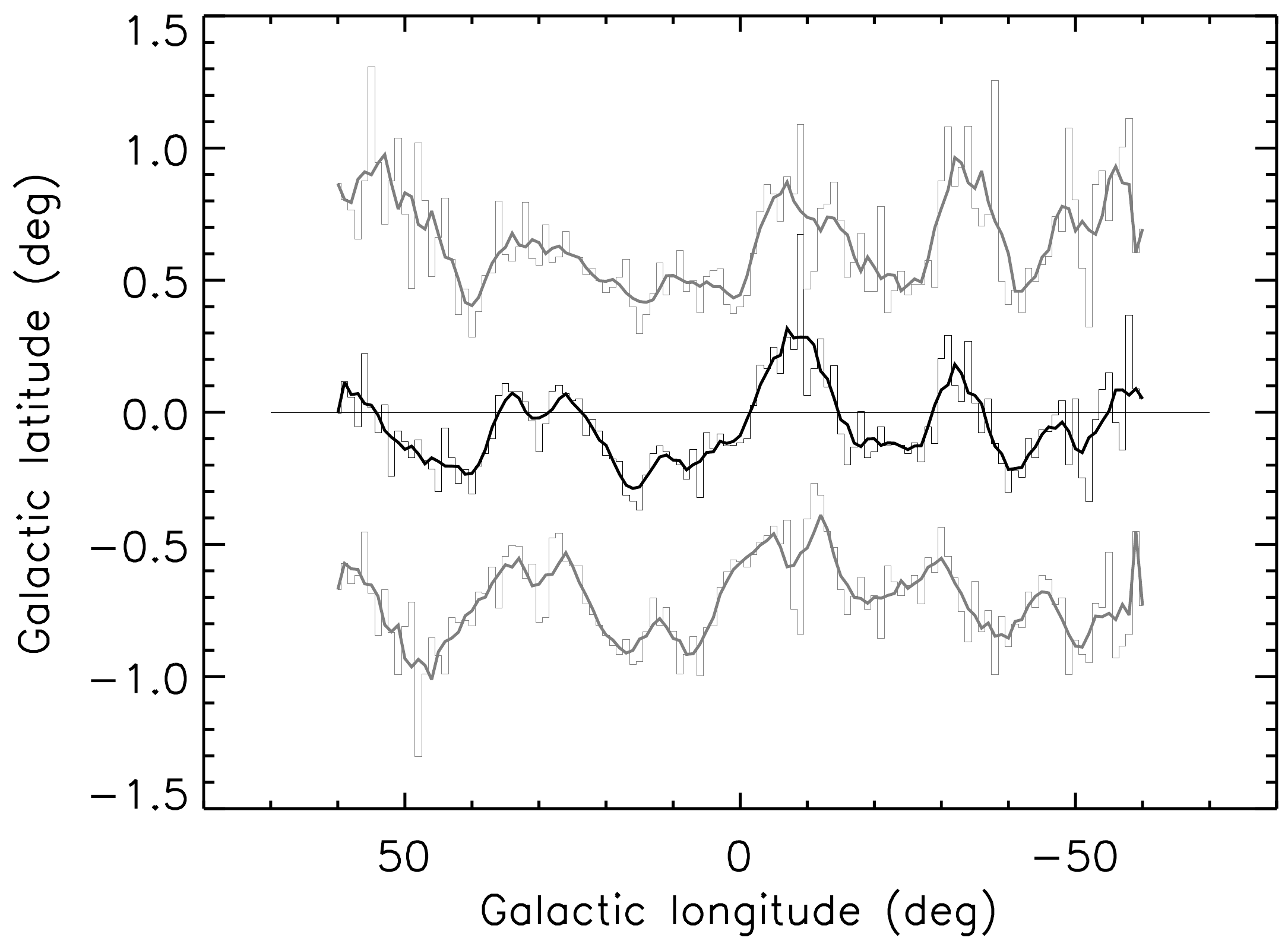}
\caption{Distribution in Galactic longitude of the latitude values for the centroid (black histogram) and the upper and lower 50\%-level values (grey histograms) of the 500-\um\ emission (see text for a detailed explanation). The thick lines represent the smoothing of the distributions with a 5\adeg -wide running boxcar.}
\label{glondistr_diff}
\end{figure}

We then compute the r.m.s. of the residual distribution and clip away all pixels whose brightness exceeds this r.m.s. by more than a factor of three (the red points in the same figure).  The polynomial order 7 was found, based on trial and error on several locations in the Plane, to be a satisfactory compromise for representing the large and medium-scale components of the diffuse emission (qualitatively $\delta \geq 10$\amin ) but not the brightest and small-scale source clusters. The remaining pixels are again fitted as a function of latitude, this time with a 5$^{th}$-order polynomial, and the maximum of the fit was assumed as the centroid of the emission.  To characterise the latitude width of the emission band we derived the latitude values where the fit reaches the 50\%\ value relative to its maximum. Again, the choice of this functional form was made after experimenting in detail over several locations in the Galactic Plane. The latitude distribution of the observed 500-\um\ flux appears heavily skewed, which dissuaded us from using a Gaussian fit; the 5$^{th}$-order polynomial appeared to be the best compromise between obtaining a reliable indication of the overall distribution peak, without being too sensitive to small-scale structures. Fig. \ref{cliptest} shows that this simple method is effective in estimating the latitude centroid of the diffuse emission (the green cross in the figure). The longitude distribution of the emission latitude centroid, as well as of the latitudes of its 50\%\ levels (above/below), are reported in fig. \ref{glondistr_diff}.

\subsection{The latitude distribution of Hi-GAL sources.}
\label{sect_distr_sources}

\begin{figure*}[t]
\includegraphics[width=1\textwidth]{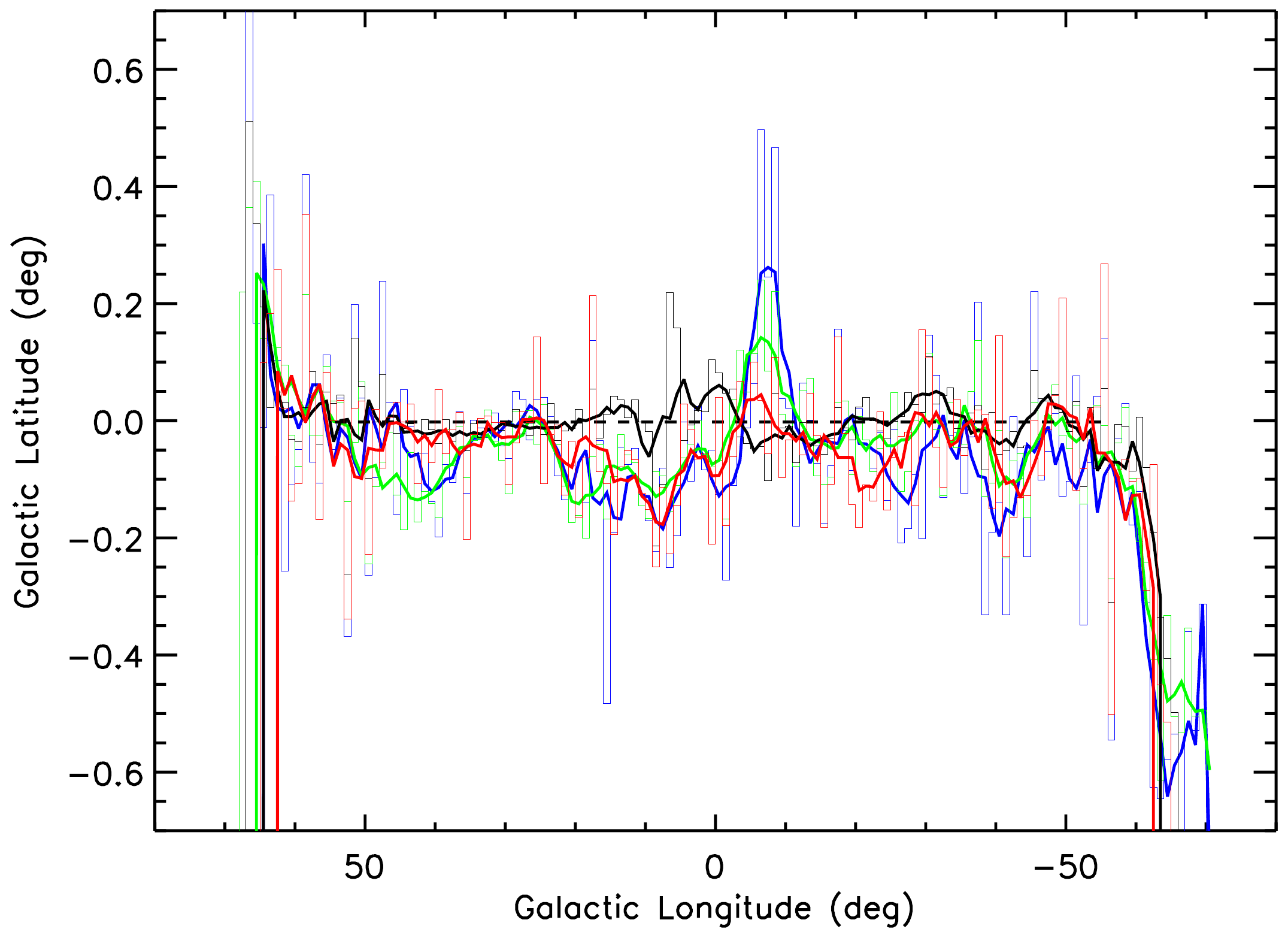}
\caption{Histograms representing the distribution of the median Galactic latitude of the Hi-GAL sources detected at 70\,\um\ (blue line) and at 250\,\um\ (green line), each with counterparts in at least two adjacent bands. The histograms are computed in 1\adeg\ longitude bins, together with the analogous distribution of entire sample of MIPSGAL 24-\um\ sources (black line); the distribution of the subsample of MIPSGAL 24\um\ sources selected by \cite{Robitaille2008} using color criteria targeted toward YSOs is shown with the red line. The thick coloured lines show the result of a 5\adeg -wide running boxcar smooth. The horizontal dashed black line reports the median latitude of all MIPSGAL 24-\um\ sources that essentially coincides with the nominal $b$=0\adeg\ midplane.}
\label{glondistr_glat}
\end{figure*}

As an additional tool to characterise the vertical distribution of the cold dust in the Galactic Plane as a function of Galactic longitude, we determine the overall latitude distribution of the compact sources found in \higal. For this, we compute the median value of the Galactic latitudes of all \higal\ sources in longitude bins of 1\adeg, limited to those sources that have a counterpart in at least three adjacent {\it Herschel} bands. This criterion is adopted to select sources with a relatively high degree of reliability (see \citealt{Elia+2015}) in the \higal\ photometric catalogues of \cite{Molinari+2015b}, resulting in nearly 100\,000 sources in the longitude range considered in the present paper. In fig. \ref{glondistr_glat} we plot these median latitude values as a function of longitude for sources with a 70-\um\ and a 250-\um\ counterpart as the blue and green histograms, respectively. By construction, all considered sources have a counterpart at 250\,\um\ (this being the central \higal\ band).  Sources with a 70-\um\ counterpart therefore form a subset (nearly 23\%) of the entire sample. To emphasise large-scale trends, we also show the same distributions smoothed with a 5-bin wide boxcar (thick lines with the same respective colour). 

It can be seen that the latitude distribution of the two classes of sources traced by {\it Herschel} is very similar over the entire longitude range plotted in fig. \ref{glondistr_glat}, implying that the 70-\um -counterpart subsample does not depart from the overall behaviour, with three exceptions. The first discrepancy is visible in the 1$^{st}$ quadrant at 50\adeg\ $\leq l \leq$ 40\adeg, where the negative bend shown by the 250-\um -counterpart sources is not followed by the 70-\um\ sources. Looking at fig. \ref{barpan}a, we note that, in this longitude range, the overall diffuse emission is clearly bending toward negative latitudes but there are discrete, large, star-forming complexes like those at $l\sim 46$\adeg\!.5 and $l\sim 43$\adeg\ at latitudes close to $b$=0\adeg. \cite{Billot+2011} carried out a detailed study of the clustering properties of \higal\ compact sources detected toward the $l$=59\adeg\ and $l$=30\adeg\ Galactic Plane fields, showing that the 70-\um\ sources have a higher degree of clustering in \hii -region complexes than sources detected at 250\,\um. It is therefore not surprising that the latitude distribution of 70-\um\ sources may be heavily biased by the presence of star-forming complexes. 
%17\adeg $\leq\ l \leq$ 20\adeg, where the boxcar-averaged 70\um\ distribution (blue thick line) shows a sudden jump to $b \sim$ 0\adeg, from the negative-$b$ trend starting at $l \sim$ 25\adeg. This is driven by the much higher jump seen in the histogram distribution, and looking at fig. \ref{barpan}b it is clear that this is entirely due to the M16 giant star fomation complex. This complex harbors by far the most prominent concentration of 70\um\ sources in the $l \sim$ 17\adeg\ region, therefore driving the median 70\um\ sources latitude to $b \sim$ 0\adeg.6 in single histogram bin. The 250\um\ sources, on the contrary, are much less concentrated, leaving the median source latitude relatively unperturbed. 
The second discrepancy between the median latitude of 70-\um\ and 250-\um\ sources is found at $-$5\adeg\ $\leq l \leq -$10\adeg\ in fig. \ref{glondistr_glat}, where the smoothed 70-\um\ distribution peaks at much higher latitudes. Again, the presence of prominent star-forming regions biases the 70-\um\ source distribution; in this case it is the NGC6334/6357 complexes, as it can be easily verified in fig. \ref{barpan}b. The third discrepancy is found at $-$23\adeg\ $\leq l \leq -$30\adeg\ in fig. \ref{glondistr_glat}, where the 70-\um\ source median latitude suddenly drops to $b \sim $ $-$0\adeg\!.3 in 4--5 histogram bins, while the 250-\um\ median stays closer to $b \sim$ 0\adeg. This discrepancy is again due to the peculiar distribution of the 70-\um\ sources in the two giant star-formation complexes visible in this longitude range: that centered around RCW 106 \citep{Mook+2004}, and the complex of HII regions around G330.986-00.433 \citep{Anderson+2014}, both of which are concentrated toward the lower boundary of the main Galactic Plane emission band.  Again, the rest of the Plane in that longitude range contributes much more in 250-\um\ sources than it does at 70\,\um. 

The reliability of Herschel source counts may be influenced by incompleteness due to confusion from extended dust emission of molecular clouds and diffuse cirrus piling up along the line of sight \citep{Molinari+2015b}. This prevents  fainter sources being detected on top of relatively stronger backgrounds so that source counts at latitudes closer to the midplane, where  the more intense background conditions are found, are depressed to an extent that is difficult to quantify. Had we to estimate and apply such correction to source counts, however, this would be higher in regions with relatively higher backgrounds, and this would amplify the results of fig. \ref{glondistr_glat} because the latitude distribution of the detected sources \textit{before the hypothetical correction} already follows the trend of the diffuse emission.

The black histogram and line in fig. \ref{glondistr_glat} report the same quantities computed from the MIPSGAL 24-\um\ point-source catalogue (Shenoy et al., priv. comm.). Sources from the MIPSGAL 24-\um\ survey represent a mix between YSOs and star-forming clumps (relatively more evolved, on average, compared to similar objects traced by the Hi-GAL source catalogues), and more evolved MS objects like post-AGB stars \citep{Carey+2009}. As a whole, the 24-\um\ MIPSGAL catalogue traces more evolved objects compared to Hi-GAL; the trend shown in fig. \ref{glondistr_glat} shows a flatter and relatively less structured behaviour with respect to Hi-GAL compact sources, with no indication of latitude modulations similar to those exhibited by the Hi-GAL compact sources. If, however, we use the sources from MIPSGAL that were selected by \cite{Robitaille2008} according to criteria that should result in a sample dominated by YSOs, we obtain (the red line in fig. \ref{glondistr_glat}) a distribution that more closely resemble the ones from Herschel. In particular, we mention the downward bends at 25\adeg\ \lapprox\ $l$ \lapprox\ 0\adeg , at $l\sim  -20$\adeg\ and $l\sim -40$\adeg .

\begin{figure}
\includegraphics[width=0.5\textwidth]{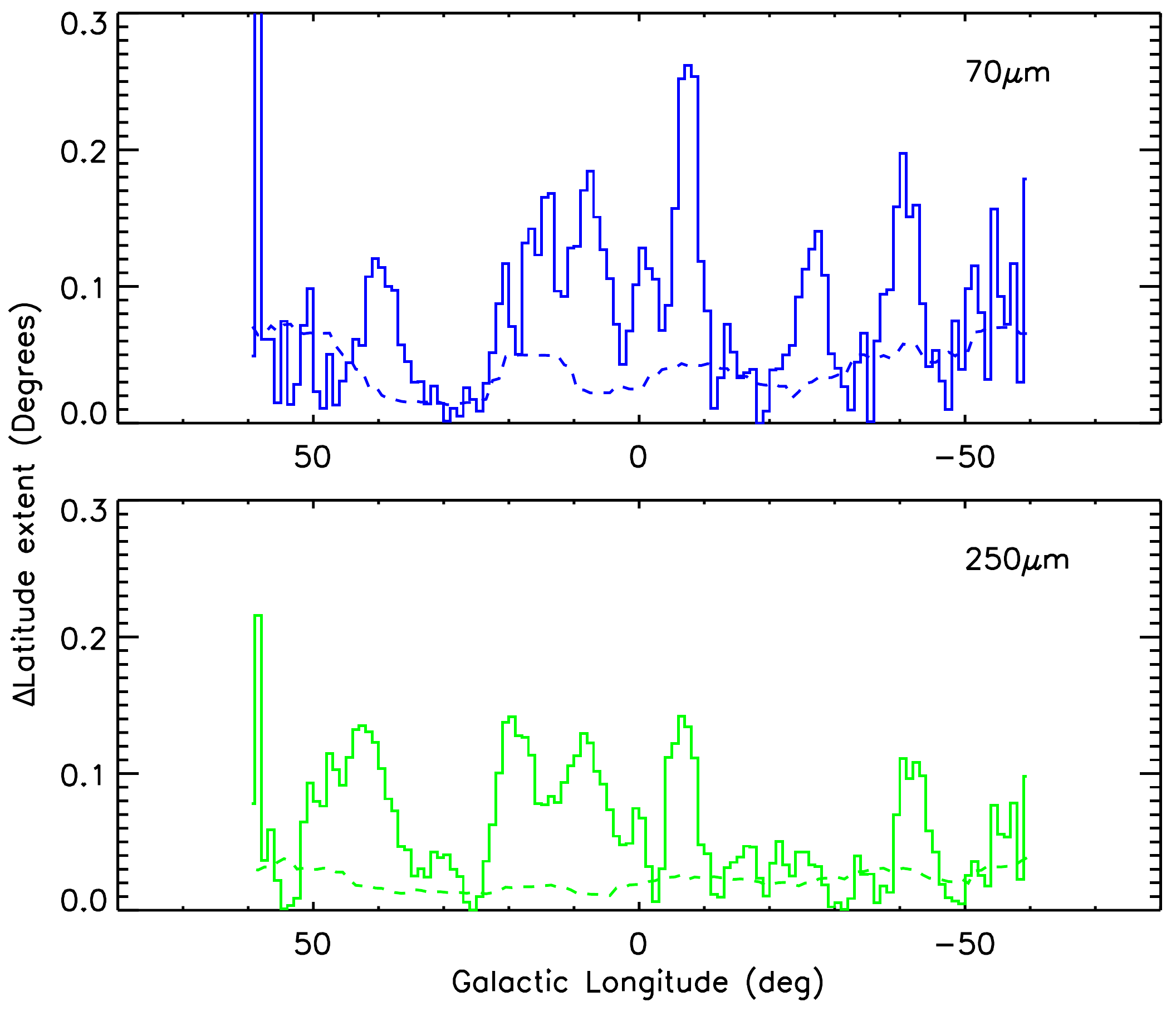}
\caption{Full-line histograms representing, as a function of longitude, the absolute value of the smoothed median source latitude (the thick lines of fig. \ref{glondistr_glat}), and the running standard deviation (dashed lines) of the residuals between the median latitude and its smoothed function (i.e., the difference between the thin and thick lines in fig. \ref{glondistr_glat}), computed using a boxcar of 10 degrees. The upper panel shows the result for the 70-\um\ sources in blue, and the lower panel shows the 250-\um\ sources in green. Note that we are using only those \higal\ sources with a counterpart in at least 3 adjacent bands.}
\label{glondistr_glat_rms}
\end{figure}

The statistical significance of the median latitude distributions of the \higal\ sources is characterised in fig. \ref{glondistr_glat_rms} for the sources with counterparts in the two \higal\ bands of fig. \ref{glondistr_glat} in the two panels. The plots show with a full line the absolute value of the smoothed median latitude of the \higal\ sources, i.e. the absolute value of the thick lines in fig. \ref{glondistr_glat}. The dashed line in fig. \ref{glondistr_glat_rms} reports the running standard deviation of the difference, as a function of the longitude, between the median source latitude and its smoothed function; in essence we computed the standard deviation of the difference between the thin and the thick lines (blue and green) in fig. \ref{glondistr_glat} in a running boxcar 10-degrees wide. The latitude distortions can be assumed significant (at least at the 1$\sigma$ level) for the longitude ranges where the dashed line is close or above the full line. In addition, we regard as a valid significance indicator the fact that the amplitude of the distortion keeps close to or above its running r.m.s. for several degrees.

%\FloatBarrier
\section{Results}

\subsection{The width of the Galactic disk}
\label{disk_width}

\begin{figure}[t]
\centering
\includegraphics[width=0.5\textwidth]{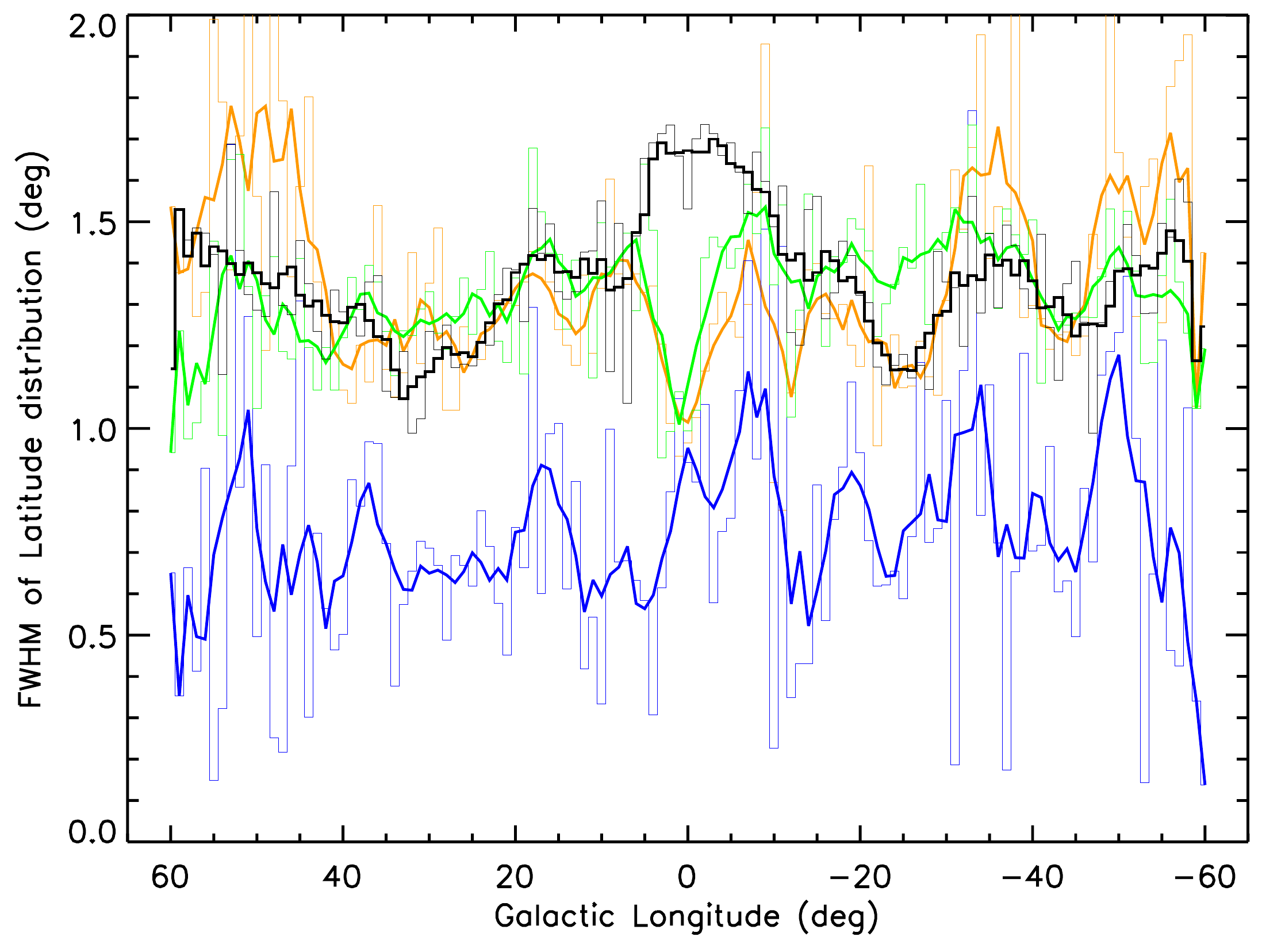}
\caption{The distribution as a function of longitude of the Full Width at Half Maximum of the latitude distribution of various quantities described in the text, averaged in 1\adeg\ longitude bins. The Hi-GAL 70-\um\ and 250-\um\ compact sources belonging to the band-merged source list are indicated with full line blue and green histograms, where the FWHMs have been determined by a Gaussian fit to the source latitude distribution in each longitude bin. The black histogram represents the distribution of the MIPSGAL 24-\um\ sources, computed in the same way as for the Hi-GAL sources. The orange histogram represents the FWHM of the Hi-GAL  500-\um\ diffuse emission obtained simply as the amplitude of the latitude band encompassed by the 50\%\ levels in fig. \ref{glondistr_diff}. For each histogram, the thick lines of the same respective colour show the distributions after smoothing with a 5\adeg -wide boxcar.}
\label{scaleheight}
\end{figure}

The amplitude of the latitude distribution of the quantities described above as a function of longitude is reported in fig. \ref{scaleheight}. The FWHM of the latitude distribution of the 70-\um\ and 250-\um\ Hi-GAL sources (blue and green lines), and the MIPSGAL 24-\um\ sources (black), are determined by fitting a Gaussian to the source latitude distribution in 1\adeg\ longitude bins. The amplitude of the 50\%-level of the Hi-GAL 500-\um\ diffuse emission (orange) is taken from fig. \ref{glondistr_diff}, and it is not the result of a Gaussian fit to the latitude distribution.

The latitude FWHM of the 250-\um\ sources and of the 500-\um\ diffuse emission range between 1\adeg\ and 1\adeg\!.5, on average, with the latter also showing peaks close to 2\adeg. The width of the Plane in these two tracers varies in a similar way with Galactic longitude. The FWHM of the MIPSGAL 24-\um\ sources follows a similar trend, with a noticeable departure toward the central molecular zone, possibly also due to the Galactic bulge; however, {\it Herschel}/SPIRE data were taken in ``bright mode" in the central 6\adeg\ around the Galactic Centre, to mitigate saturation at the expense of sensitivity, and this may be the origin of the dip in the FWHM of the latitude distributions of both Hi-GAL 250-\um\ sources and 500-\um\ diffuse emission. A similar FWHM of the Plane is also reported by \cite{Beuther2012} for the GLIMPSE red sources in the \cite{Robitaille2008} sample. 

The distribution of the Hi-GAL 70-\um\ sources traces a much thinner Plane, with the FWHM showing large variations between $\sim$0\adeg\!.3 and 1\adeg\!.3, with a mean value of 0\adeg\!.75 for $-$60\adeg $\leq l \leq$+60\adeg. This is slightly larger than the 0\adeg\!.6 value reported for the ATLASGAL 870-\um\ sources by \cite{Beuther2012}. The amplitude of the variations in width for the 70-\um\ sources is larger than for the 250-\um\ sources, and can be explained by the fact that sources with a 70-\um\ counterpart trace the prominent star-forming and \hii -region complexes more closely \citep{Billot+2011} and therefore can generate strong local departure from larger-scale longitude trends. 

\begin{figure}[t]
\centering
\includegraphics[width=0.5\textwidth]{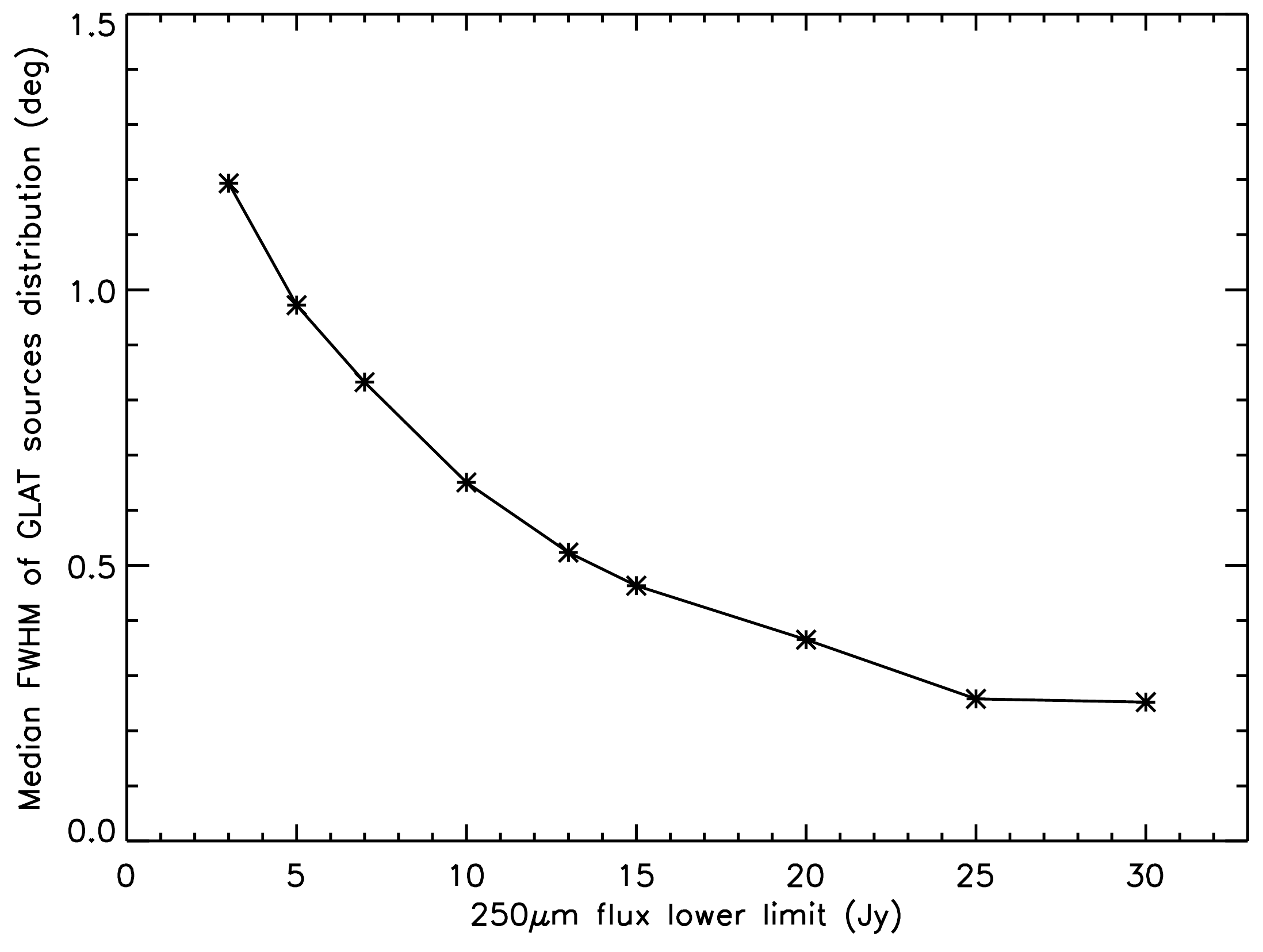}
\caption{Median values, over the whole longitude range, of the FWHM of the latitude distribution, in bins of 1\adeg, of the 250-\um\ \higal\ sources with integrated flux above a certain limit, as a function of said limit.}
\label{scaleheight_250}
\end{figure}

It may seem puzzling that the \higal\ 250-\um\ sources do not show the same latitude distribution as the ATLASGAL 870-\um\ sources \citep{Beuther2012}, while the \higal\ 70-\um\ sources do. APEX/LABOCA at 870\,\um\ has the same spatial resolution of {\it Herschel}/SPIRE at 250\,\um, and the wavelengths are both in the Rayleigh-Jeans section of the SED of cold dust, so that one would naively expect the two samples to exhibit a similar behaviour in their statistical distributions. The discrepancy is due to the significantly different sensitivities of the two instruments, coupled with the different completeness limits that can be reached on and off the midplane at 250\,\um. In the most recent version of the ATLASGAL source catalogue, \cite{Csengeri+2014} report the peak of the 870-\um\ integrated flux distribution at $\sim 0.6$ Jy. Extrapolating to 250\,\um\ assuming $\beta=1.5$, this would correspond to a flux of $\sim$47\,Jy. On the other hand, the peak of the 250-\um\ source integrated flux distribution is at $\sim 3$\,Jy \citep{Molinari+2015b}, showing that {\it Herschel} is, as expected, at least 10 times more sensitive than current top-notch ground-based submillimetre surveys. The ATLASGAL sources therefore represent the highest-flux fraction of the \higal\ compact sources, suggesting that the different latitude distributions of the two samples is due to a bias from the different ranges of fluxes sampled. To verify this, we analyse again the latitude distribution of the \higal\ 250-\um\ sources, using only those above a certain minimum integrated flux. For each choice of this minimum flux, we fit with a Gaussian the latitude distribution of the sources in 1\adeg longitude bins and compute the median of the FWHMs between $-60$\adeg and +60\adeg. Fig. \ref{scaleheight_250} reports the resulting median FWHM of the 250-\um\ sources as a function of the 250-\um\ minimum flux adopted, and proves that the smaller FWHM of the ATLASGAL source latitudes, with respect to \higal\ 250-\um\ sources, is likely to be a sensitivity effect. 

\begin{figure}[t]
\centering
\includegraphics[width=0.5\textwidth]{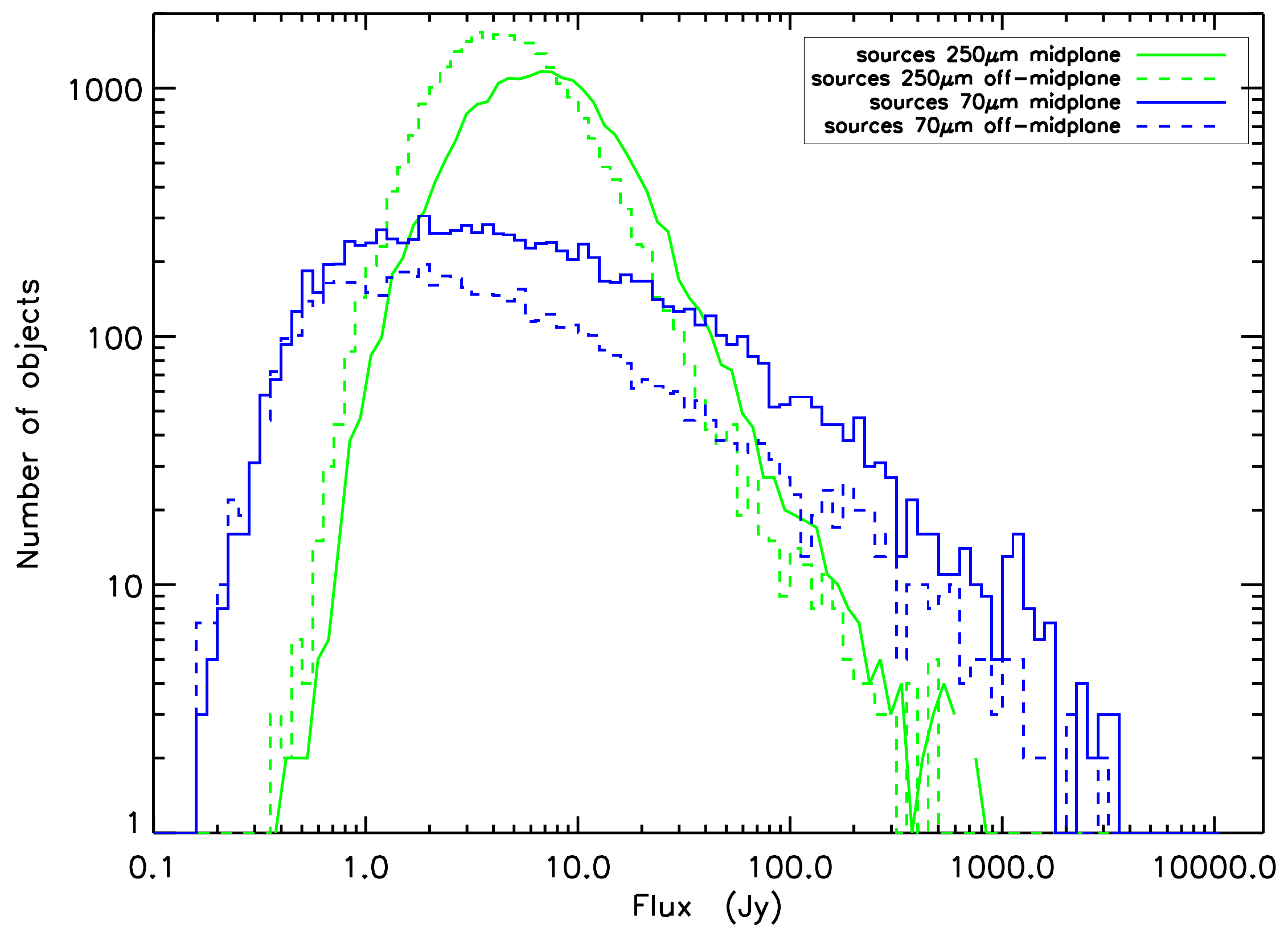}
\caption{Flux distribution for \higal\ sources at 250\,\um\ (green lines) and at 70\,\um\ (blue lines), for sources with latitudes $|b| \leq$ 0\adeg .375 (full lines) and $|b| >$0\adeg\!.375 (dashed lines). The value of 0\adeg\!.375 corresponds to half the mean (over longitude) FWHM of the latitude distribution of the 70-\um\ sources. Notice the shift in of the 250-\um\ source distributions for different latitude cuts.}
\label{histoflux_midplane}
\end{figure}

The reason for the different latitude distribution of SPIRE 250-\um\ sources compared to ATLASGAL sources has then to be sought in the population of relatively faint \higal\ sources. Cirrus noise considerably changes as a function of latitude for any given longitude in the inner Galaxy.  As a consequence, the flux completeness limits for extracted sources vary too.  As shown in fig. \ref{histoflux_midplane}, the distribution of 250-\um\ fluxes for sources with $|b| \leq 0$\adeg .375 (half the mean FWHM of the 70-\um\ source latitude distribution as determined from fig. \ref{scaleheight}), peaks at about $\sim$7\,Jy (full green line), while sources at larger latitudes peak at $\sim$4\,Jy (dashed green line); the distributions show a clear shift. This means that in the brightest cirrus regions closer to the Galactic midplane we are less efficient (due to confusion noise) in picking up relatively fainter sources.  This causes a deficit in the 250-\um\ source counts in the central latitudes, leading to an artificially shallower distribution. No such effect is present at 70\,\um\ (the blue lines in fig. \ref{histoflux_midplane}).

%This also shows, on the other hand, that the highest flux clumps are more concentrated toward the local location of the midplane defined not by $b=0$\adeg, but by the local median source latitude distribution as shown in fig. \ref{glondistr_glat}. This will be discussed in greater detail in the rest of the paper.

%It is interesting how the FWHM of the latitude of the \higal\ 70\um\ sources matches that of ATLASGAL and also that of the \higal\ 250\um\ sources with integrated fluxes above $\sim$12 Jy (see fig. \ref{scaleheight_250}). This suggests that the sensitivity of PACS at 70\um\ is intrinsically lower than SPIRE at 250\um, as expected given the predicted instrumental sensitivities \textbf{(***references from instrument manuals***)}. If we select only the \higal\ sources at 250\um\ that have a positionally matching counterpart at 70\um\ (within half of the SPIRE beam at 250\um, or $\sim$6\asec) we find that the fluxes at the two bands are related by a power law relationship of the type $Log(F_{250})\sim 0.71 + 0.71 Log(F_{70})$. Using this relationship a 250\um\ flux of 9 Jy, corresponding the lower flux limit where \higal\ 250\um\ sources have the same latitude FWHM of the \higal\ 70\um\ sources and of ATLASGAL sources, would correspond to a 70\um\ flux of 2.2 Jy; this is very similar to the flux where the histogram of the \higal\ 70\um\ source fluxes peak \citep{Molinari+2015b}. We conclude that \higal\ 70\um\ sources trace the population of \higal\ 250\um\ sources for fluxes $F_{250} \geq$10 Jy, and most likely the population of ATLASGAL 870\um\ objects.

\subsection{Large-scale latitude distortions of the dusty Galactic disk}
\label{distortions}

In this section we analyse in detail the latitude distribution of compact sources and diffuse emission from \higal. We will make extensive use of figures \ref{barpan}, \ref{glondistr_diff}, \ref{glondistr_glat} and \ref{glondistr_glat_rms}.

\subsubsection{The First Quadrant}
\label{dist_1}

Fig. \ref{barpan}a shows the emission map at 500\,\um\ in the first Quadrant starting from $l\sim$67\adeg, where the Hi-GAL coverage starts to climb up in latitude to follow the Galactic Warp. In a popular representation of the four-arm Milky Way Galaxy, the region 60\adeg $\leq l \leq $50\adeg\ is dominated by the Local Arm; many noticeable star-forming complexes in this region are located at heliocentric distances of a few kpc at most (e.g., the Vul OB1 complex, see \citealt{Billot+2010}). As soon as we see the distribution of the bulk of the emission coming down from the warped disk onto the $b=0$\adeg\ midplane, at around $l=60$\adeg, the latitude distribution of the emission appears to be overall centred around the midplane. As we move to smaller longitudes we see that, starting from $l\sim 50$\adeg, approximately corresponding to the tangent point of the Sagittarius Arm, the overall distribution of the 500-\um\ emission, neglecting local oscillations at the sub-degree scale, begins to be more and more shifted towards lower latitudes. On a purely visual level, it is immediately apparent that the majority of the emission at 50\adeg $\geq l \geq$ 36\adeg\ in fig. \ref{barpan}a is mostly concentrated below $b=0$\adeg.  The curved shaded area superimposed on the map is meant to guide the eye along this downward bend of the overall emission that, after reaching a minimum around $l\sim$ 40\adeg, comes up to rejoin the $b=0$\adeg\ midplane at around $l\sim 36$\adeg. The distribution remains centred at $b\sim$ 0\adeg\ as we proceed to smaller longitudes, where the Scutum-Centaurus Arm starts to dominate, through the W43-G29.9 complex and toward the $l\sim$ 25\adeg\ region. The negative bend in the latitude distribution of the emission seems therefore confined to the portion of Plane dominated by the Sagittarius Arm. This behaviour for 50\adeg $\leq l \leq$ 36\adeg\ is quantitatively confirmed in fig. \ref{glondistr_diff}, where the 500-\um\ emission centroid is indeed staying consistently below $b=0$\adeg\ for 50\adeg $\leq l \leq$ 36\adeg , reaching a minimum value of $b \sim -0$\adeg\!.25. This displacement appears to be significant judging by the dispersion of the centroid histogram (thin histogram) with respect to its boxcar running average (thick black line) in fig. \ref{glondistr_diff}. 

The same can be seen in the latitude distribution of the compact sources in fig. \ref{glondistr_glat}. Here, the amplitude of the latitude dip is relatively smaller (less than 0\adeg\!.2), but it is visible both for the 70- and the 250-\um\ sources. Fig. \ref{glondistr_glat_rms} shows that this bend is significant to about the 2-$\sigma$ level (blue and green dashed lines are close or well above the full lines). We again note that although the effect is strictly speaking marginal, we are looking at degrees-wide longitude sections where the dashed line stays above the full line.
It is interesting to note here that the latitude distribution of the 24-\um\ compact sources from MIPSGAL does not follow the dip but stays remarkably flat and close to $b=0$\adeg, even if the \cite{Robitaille2008} YSO-dominated 24\um\ sample is considered.

\subsubsection{The Galactic Bar}
\label{dist_bar}

Fig. \ref{barpan}b shows in two subpanels a much larger longitude range from $l\sim 27$\adeg, through the Galactic Centre, to $l\sim 328$\adeg. It is dominated by the Scutum-Centaurus arm from its tangent point in the 1$^{st}$ quadrant, the entire Galactic Bar and Central Molecular Zone, and the Norma Arm and its tangent point in the 4$^{th}$ quadrant. At around $l\sim 25$\adeg, the emission distribution that has remained centred on $b\sim 0$\adeg\ since  $l \sim 36$\adeg\  (see previous section) starts again to bend slowly downward. For longitudes smaller than 20\adeg, the emission is consistently found at negative latitudes; again, the shaded area on the figure is useful to guide the eye along this bend. This is the near side of the Galactic Bar. The emission band starts to climb up again below $l \sim$ 5\adeg, goes across the midplane at the Galactic Centre, and up to positive latitudes in the fourth Quadrant. The lower panel of fig. \ref{barpan}b clearly shows that the barycenter of the emission appears consistently above $b\sim 0$\adeg\ for longitudes 358\adeg$\geq l \geq$ 345\adeg\ ( $-$2\adeg$\geq l \geq -$15\adeg), corresponding to the far side of the Galactic Bar, approaching again $b\sim 0$\adeg\ roughly at the location of the far tip of the Bar. Beyond this location we see again that the emission barycenter heads to negative values until the tangent point of the Norma Arm is reached, at around $l\sim 330$\adeg\ (or $l\sim -30$\adeg.).

As in the previous section, the visual impression from the panoramic images is quantitatively verified using the latitude distribution of the 500-\um\ emission barycenter in fig. \ref{glondistr_diff}, as well as the distribution of the median latitude of the compact sources in fig. \ref{glondistr_glat}. The figures confirm that the Plane, as traced by both the diffuse 500-\um\ emission and by the 70- and 250-\um\ compact sources, stays around $b\sim$0\adeg\ for 35\adeg$\leq l \leq$ 25\adeg, making a negative latitude bend all the way across the Galactic Centre, coming up to positive latitudes after that until $l \sim -15$\adeg. We believe the evidence is robust because the above trend is also consistently shown by the distribution of the 50\%\ levels of the 500-\um\ emission (the grey lines in fig. \ref{glondistr_diff}), as well as confirmed, with similar shape and magnitude, by the 70-\um\ and the 250-\um\  compact-source distribution. The only two locations where the distributions of the 70-\um\ and 250-\um\ sources differ is at $l\sim 18$\adeg, where the 70-\um\ distribution is well above the 250-\um\ one and goes back to $b\sim$ 0\adeg), and at $l\sim -6$\adeg, where the upward climb of the 70-\um\ sources is much more marked. These two occurrences are due to the local bias in the median estimate of the 70-\um\ source latitude induced by the large M17 and NGC6334/6357 star-forming complexes. 
These distortions are significant over the range +25\adeg $\geq l \geq -10$\adeg\  at the 7-$\sigma$ level for the 250-\um\ sources and at 4--5$\sigma$ for the 70-\um\ sources (from fig. \ref{glondistr_glat_rms}).  For the latter, the r.m.s. of the latitude distribution is larger than the amplitude of the distortion between 20\adeg\ and 15\adeg, due to the peculiar latitudes of the M16/M17 complexes with respect to the local median.  Again, the significance should be judged not only by pure $\sigma$ levels over the r.m.s., but also based on the persistence of the significance over an area a few degrees wide.

There is an additional downward bend that can be visually identified at 345\adeg $\geq l \geq$\,330\adeg\ in fig. \ref{barpan}b that is also visible in fig. \ref{glondistr_diff} at $-15$\adeg $\geq l \geq -$30\adeg\ , although with a lower magnitude compared to the region of the Galactic Bar. The distribution of the 70-\um\ sources shows the same trend, but the 250-\um\ one does not and remains centered at $b\sim $ 0\adeg. We regard the Plane distortion in this region as less certain. Again, we note how the latitude distribution of the 24-\um\ compact sources from the entire MIPSGAL sample does not follow any of the patterns exhibited by the diffuse emission or the compact-source distribution in Hi-GAL. The situation is different for the YSO-dominated sample of \cite{Robitaille2008} that instead exhibits the downward bend between 20\adeg and 0\adeg.

\subsubsection{The Fourth Quadrant}

The appearance of the Plane in the diffuse 500-\um\ emission in the rest of the 4$^{th}$ quadrant is shown in fig. \ref{barpan}c, where the band of emission is mostly centred at $b\sim$ 0\adeg. There is a downward bend at 325\adeg $\geq l \geq$ 315\adeg\ ($-$35\adeg $\geq l \geq -$45\adeg) that is visible in fig. \ref{glondistr_diff} and in the compact-source distribution in fig. \ref{glondistr_glat}. The overall emission along the Plane in this region is more inhomogeneous with respect to the Bar and the 1$^{st}$ quadrant, with gaps of relatively faint emission. The significance of the distorsion patterns seems however reliable.
% This lessens the reliability of the estimate of the latitude centroid in the diffuse emission, as the fit is poorly constrained, and decreases the statistical significance of the median latitude of compact sources because their number drops in these areas. 
Moving further away from the inner Galaxy, at $l\leq $ 310\adeg\ (or $l\leq -$50\adeg) the 500-\um\ Plane emission appears much more widely spread in the vertical direction, which explains the relatively higher level of channel-to-channel noise in figs \ref{glondistr_diff} and \ref{glondistr_glat}.

\subsection{Characterising biases for the large-scale distortions.}

\subsubsection{Distance effects}

The position of the Sun $\sim$27 pc above the nominal $b=0$\adeg\ Plane has to be considered as one of the possible reasons why the latitude distribution of Galactic sources retrieved in Hi-GAL or in ATLASGAL is generally peaked to slightly negative latitudes \citep{Molinari+2015b, Beuther2012}. This effect would, however, also generate distance-dependent distortions, in that closer objects would appear at more negative latitudes than more distant objects. 
To check if this may be the cause of the observed distortions, we made a preliminary investigation using the subset of \higal\ sources for which a heliocentric distance estimate is available and that will be presented by \cite{Elia+2015}. 
%This subset comprises at the moment almost 60\,000 sources which exhibits a \higal\ detection in at least three adjacent Herschel bands, representing therefore around 20\%\ of the total objects in the \higal\ 250\um\ photometric catalog released by \cite{Molinari+2015b}.

The longitude range of the Galactic Plane covered in the release of \cite{Molinari+2015b} was divided in bins of 1\adeg, and for each bin we computed the distance distribution of the objects along the line of sight in bins of 1 kpc. The distance at which the distribution peaks is likely to provide the major contribution, in terms of source numbers, to the determination of the median source Galactic latitude reported in fig. \ref{glondistr_glat}. We find a location at $d\sim 7-10$\, kpc for the majority of the sources in the 35\adeg$\leq l \leq$50\adeg\ longitude range, $d\sim 10-13$\,kpc for the 12\adeg$\leq l \leq$ 20\adeg\ range (we do not have distance determinations for sources closer to the Galactic Centre at the moment), and $d\sim 1.5-3$\, kpc for the 340\adeg$\leq l \leq$ 350\adeg\ range. None of these preliminary findings agrees with the expectations for a latitude distortion induced by distance effects. 

We therefore conclude that the apparent distortions are real. A more thorough analysis will be needed \citep{Elia+2015} to verify in detail to what extent the distortions are an overall property of the Galactic disk or are confined to specific arm sections.

\subsubsection{Source completeness biases in 24\um\ catalogues}

We investigate here the possibility that the shape of the source latitude distributions as a function of longitude as reported in fig. \ref{glondistr_glat} for the entire sample of MIPSGAL 24\um\ sources (the black line), may be biased by incompleteness in source catalogues due to confusion from extended emission or extinction.

\begin{figure}[t]
\includegraphics[width=0.5\textwidth]{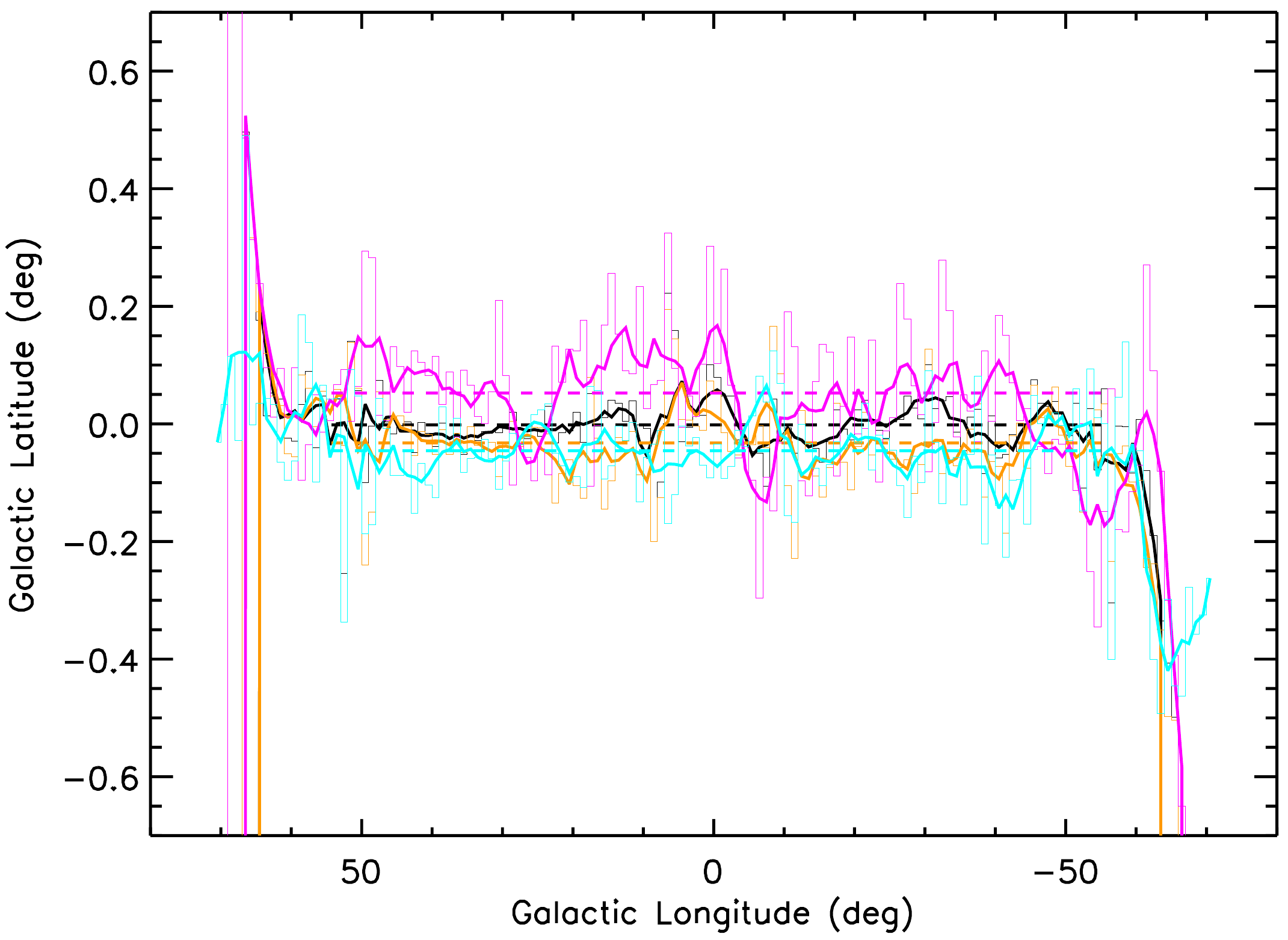}
\caption{Same as fig. \ref{glondistr_glat} for the distribution of MIPSGAL 24 \um\ sources only. The black line is reported straight from fig. \ref{glondistr_glat} for comparison. The orange line represents the distribution again from the Shenoy et al. (in prep) catalogue but for sources with fluxes above 3.5mJy. The cyan line represents the distribution from the WISE 22\um\ source catalogue with flux above 6 mJy. The magenta line instead represents the distribution of the 24 \um\ sources from the catalogue in \cite{GH2015} that are above the respective 90\% completeness limit.}
\label{glondistr_glat_mgal}
\end{figure}

The situation is different for the distribution of MIPSGAL 24 \um\ sources (the black line in fig. \ref{glondistr_glat}); here a possible limitation to the detection of faint sources is extinction by dust. In this case the effect should be higher where the dust column is higher along the Plane, as traced by the 500 \um\ emission (see figures \ref{barpan} and \ref{glondistr_diff}). The fact that, contrary to what happens for the \higal\ 70 and 250 \um\ sources, the distribution of the 24 \um\ sources \textit{does not show} the same oscillations in latitude centroid as a function of longitude exhibited by the 500 \um\ diffuse emission (the black line in fig. \ref{glondistr_diff}), could then be due to dust extinction that depresses the 24 \um\ source counts on the peaks of the 500 \um\ emission. A way to mitigate this bias effect is to compute the distribution of 24 \um\ sources only for objects with fluxes above the completeness limit, so that each latitude bin is equally treated providing a less biased median latitude estimate. There is currently no flux completeness limit for the yet unpublished Shenoy et al. catalogue; however, a simple histogram of the fluxes shows a peak around $\sim$ 3 mJy. Since the turn-down of fluxes histograms from photometric catalogues is typically very close to the completeness limit (e.g. \citealt{Molinari+2015b} for the \higal\ catalogues)  we tentatively assume 3 mJy as an average value for 24 \um\ completeness in the Shenoy et al. catalogues. Carrying out the same analysis of fig. \ref{glondistr_glat} only for sources with flux above 3mJy results in the orange line in fig. \ref{glondistr_glat_mgal}; this differs somewhat from the black line, but shows again very limited latitude variations around its respective midplane (the orange dashed line), and certainly does not resemble the distribution of the \higal\ 70 and 250 \um\ sources. The same conclusion is reached if we use the WISE 22\um\ source catalogue; this time we use only sources with flux above $\sim$ 6 mJy, roughly corresponding to the peak of the flux distribution of WISE sources between -60\adeg\ $\leq l \leq$ 60\adeg\ and |$b$|$\leq$ 1, and using the same processing as above we obtain the cyan line in fig. \ref{glondistr_glat_mgal}. Again the distribution is much flatter around its respective midplane (cyan dashed line) than that of the \higal\ sources.

As an additional check, we also used the very recent 24 \um\ MIPSGAL catalogue of \cite{GH2015} that contains high-reliability objects and very conveniently includes for each source an estimate of the local 90\%\ flux completeness value. We carry out the same analysis again for the \cite{GH2015} catalogue only using sources above the completeness and the result is shown by the magenta line in fig. \ref{glondistr_glat_mgal}. The latitude variations of the magenta line with respect to its median-determined midplane (the magenta dashed horizontal line) are still substantially limited, and again completely differ from the \higal\ sources. There are differences between the latitude distribution of 24 \um\  sources between Shenoy et al. and \cite{GH2015}, but this is beyond the scope of the present paper that is focused on Herschel data. 

We conclude that completeness or extinction biases over the entire sample of detected 24\um\ sources cannot explain the intrinsically different shapes of the longitude distribution of sources' latitude when the entire 24\um\ catalogues are considered. It is only when the 24\um\ catalogues are filtered with criteria that preferentially select YSOs that the the 24\um\ sources distribution more closely approaches the distribution of the Herschel sources.

\section{Tracer-dependent disk distortions: evidence for an external agent.}

Our results confirm the findings of the studies mentioned earlier (see \S\ref{intro}) with much higher reliability and statistical significance, but it is noteworthy that those previous investigations were basically successful in identifying the large-scale latitude modulations in the distribution of gas and star-forming tracers using relatively small or limited data sets. 

One fact to recall at this point is the different appearance in the latitude distribution of the Galactic disk as a function of the tracer. For instance, the HI elliptical disk distribution in the CMZ is tilted by $\sim$24\adeg~\citep{McClure+2012}, i.e significantly more than the inclination angle of $\sim$ 1\adeg\!.27 that we infer from the median longitude distribution of the latitude centroid of the {\it Herschel} 500-\um\ emission via a simple linear fit to the black line in fig. \ref{glondistr_diff} in the range 17\adeg\!.5 $\gtrsim l \gtrsim -10$\adeg .

The different morphology in the latitude distribution of the Galactic disk as a function of the tracer can provide clues concerning the possible agents responsible for driving the perturbations. \cite{Alfaro+1992} report modulations along the Sagittarius-Carina arm, not only in the star-formation and gas tracers but also in the young stellar populations traced by young clusters.  These deviations have a maximum semi-amplitude of about 50\, pc, assuming that the clusters are indeed distributed along the Sag-Car arm between $l$=20\adeg\ and $l$=280\adeg\ (going through the Galactic Centre). At the heliocentric distance of the Sag-Car arm toward the $l$=300\adeg\ line of sight, 50\, pc correspond to an angular amplitude of more than 1\adeg\!.8. Such large amplitude modulations are not measured in the latitude distribution of objects in recent near/mid-IR Galactic Plane Surveys. \cite{Benjamin+2005} does not report any distortion of the stellar component traced by the Spitzer/GLIMPSE survey, showing a mean latitude source distribution close to $b$=0\adeg. The mean latitude distribution of the sources detected in the Spitzer/MIPSGAL 24-\um\ survey, represented by the black histogram and full thick line in fig. \ref{glondistr_glat}, tends to depart from a perfect flat distribution, showing some distortions that seem uncorrelated and of much lower amplitude than in the {\it Herschel} 70-\um\ and 250-\um\ sources; certainly, nothing similar to what proposed by \cite{Alfaro+1992} is revealed. We argue that the limited number of clusters (28) used in \cite{Alfaro+1992} is not representative of the stellar component of the disk in the Sag-Car arm. To summarize, we are presented with a complex scenario where the latitude distribution of the Galactic disk behaves differently depending on the tracer considered. Latitude modulations as a function of longitude are observed in the atomic and relatively less dense phase of the ISM traced by the HI 21-cm line. The same modulations, with a lower amplitude, are revealed in the denser ISM phase traced both by the large-scale diffuse cold dust emission, and by the distribution of the star-forming clumps revealed at $\lambda \geq$ 70 \um\ in the {\it Herschel} Hi-GAL survey. No evidence of such modulations are found in the point sources detected by the Spitzer MIPSGAL and, to the best of our knowledge, GLIMPSE surveys, when the entire source samples are considered. If instead we consider the subsample selected based on color criteria targeted toward YSOs as in \cite{Robitaille2008}, the distribution exhibits some of the characteristics of the Herschel sources distributions.

It is difficult to interpret this picture by invoking gravitational perturbations induced by the Milky Way satellites, as gravity would indifferently act both on gas and stars, while the distortion here is only seen in diffuse gas and dust and the star-forming component of the Galaxy (but see \citealt{Carlin+2013}). Likewise, any dynamical perturbation induced by non-axisymmetric agents like the Galactic Bar should act on all disk components as well. A plausible driving agent has to dynamically interact with the diffuse, gas and dust content of the disk but have negligible cross-section with the stellar component.  We propose that such an agent may be constituted by incoming flows of relatively low-density diffuse material from the Galactic halo accreting onto the Galactic disk and/or by Galactic fountain gas. 

\subsection{Interaction between gas flows and the Galactic disk}

While a detailed model of the interaction between the Galactic disk and incident extraplanar gas flows is beyond the scope of this paper, it is instructive to verify the plausibility of this interpretation. Approximating the Galaxy as a thin disk of uniform surface density $\sigma _0$, the vertical gravitational field at a height $z$ above (or below) the plane can be written as 

\begin{equation}
%{{dF_z}\over{dz}}\approx -2\pi G \sigma _0 \left[1-{{z/R_0}\over{\sqrt{1+(z/R_0)^2}}}\right]
g\approx -2\pi G \sigma _0 \left[1-{{z/R_0}\over{\sqrt{1+(z/R_0)^2}}}\right]
\label{field}
\end{equation}

In our specific case we take 0\adeg .2 as a representative distortion amplitude from fig. \ref{glondistr_glat}. The linear vertical amplitude $z$ will of course depend on where in the inner Galaxy the distortion is located, but even at the location of the CMZ this would correspond to about 30 pc; this is negligible with respect to whatever choice we make for the Galaxy radius $R_ 0$. In other words $z/R_0 \ll 1$, so that the term in square brackets in eq. \ref{field} is $\sim$1. A fraction of the disk of unit area undergoing a vertical distortion above the plane, with amplitude $z$ much smaller than the Galaxy radius, will then experience a downward force 

\begin{equation}
F_z\approx -2\pi G \sigma _0 ^2 
\label{fz}
\end{equation}
in the above we neglect that the field in eq. \ref{field} includes the contribution from the unit area undergoing the distortion. Assuming that the distortion is a stationary feature, this force will have to equal the force exerted by an incoming flow of gaseous material at velocity $v_f$, that we can write as 

\begin{equation}
F_{f}=\dot m_f v_f
\end{equation}
which, for a unit area flow column, can be rewritten as

\begin{equation}
F_{f}=\rho _f v_f ^2
\label{mom}
\end{equation}
At equilibrium $F _z = F_f$, so that the required volume density of the incoming flow is

\begin{equation}
\rho_f={{2 \pi G \sigma _0 ^2}\over{v_f ^2}}
\label{flowdens}
\end{equation}

For a typical average velocity of High-Velocity Clouds (REF) of $v_f \sim$ 200 \kms , and a mean Galaxy surface density of $\sigma _0 \sim$0.015 \gcmtwo\ \citep{Bovy_Rix2013}, we obtain $\rho _f \sim$ 0.077 \cmthree. These volume densities, as approximate as they are in this very simplified treatment, are  compatible with measurements of extraplanar gas in Intermediate and High-Velocity Clouds (e.g., \citealt{Putman+2012} or \citealt{SB1991}, see below next paragraph).

Such flows would provide an important large-scale momentum input for ordered flows in the disk or to drive turbulence, two of the main mechanisms invoked for the origin of dense ISM clouds and filaments where supercritical conditions for star formation exist (\citealt{Molinari+2014} and references therein).

\subsection{Galactic fountain or accretion from halo flows ?}

 The large-scale ordered appearance of the distortions revealed by {\it Herschel} would suggest that these features are stable over fractions of the orbital period. One can ask how this time scale compares with that of such dynamical processes as produced by ``Galactic fountains''. To drive a Galactic fountain, a series of massive stars from OB associations have to generate supernovae to guarantee that sufficient momentum is available to create a large-scale fountain. Assuming a Salpeter IMF, \cite{McCrayKafatos1987} estimated that this phase lasts 5$\times 10^7$ yrs, on average. When the expelled gas falls back it will impact the Galactic disk that is locally rotating with a tangential velocity of $\sim$220\,\kms\ basically everywhere in the inner disk.
 
Direct evidence of this gas mainly comes from UV spectroscopy that \cite{SB1991} modelled with a $n_{\rm H} \sim 10^{-2}$\,cm$^{-3}$ medium accreting onto the Galactic disk over linear extents exceeding a kiloparsec. Assuming that the infalling fountain lasts the same amount of time as the series of supernova explosions, the spatial extent of the interface among the two flows will be in excess of 10\,kpc, comparable to the linear extent that we can reasonably project from the angular extent of the observed disk bending (see e.g. \citealt{Fraternali+2015}). The Galactic fountain hypothesis is therefore compatible with the extent of the observed features from a timescale viewpoint.
%where the linear extent of the oscillation pattern like one that is currently observed corresponds to the complete duration of the interaction with a fountain driven by the final evolutionary phases of an OB association. 

% the point of view of momentum conservation, we have to consider the interaction of the ISM in the Galactic disk, with average densities between 1 and 10 \cmthree\ \citep{Pineda2013}, and a more tenuous infalling fountain gas. Direct evidence of this gas mainly comes from UV spectroscopy that \cite{SB1991} modelled with a $n_{\rm H} \sim 10^{-2}$\,cm$^{-3}$ medium accreting onto the Galactic disk over linear extents exceeding a kiloparsec. In order to transfer sufficient momentum to the denser disk ISM, moving with a rotation velocity of the order of 200\,\kms, the fountain gas should be falling back onto the disk at a speed at least an order of magnitude higher to cause the observed bending. No present observation supports the existence of such extremely high velocity infalling gas.

%\subsection{Accretion from halo flows.}

Accretion onto galaxies from halo gas that is intergalactic in origin is commonly invoked to solve the gas depletion problem in star-forming galaxies. At an average Galactic star-formation rate of $\sim 1$\msunyr, the Milky Way would exhaust its current reservoir of molecular gas in about 2$\times  10^9$ years, clearly requiring replenishment of fresh ISM material from the surrounding halo; we refer the reader to the recent reviews of \cite{Sancisi+2008} and \cite{Putman+2012}. \cite{Frater+2012} suggest that the very existence of a star-formation Schmidt-Kennicutt law requires continuous fuelling of gas, either from gas stripping from orbiting dwarf galaxies or accretion from a diffuse halo. \cite{Peek2009} simulates gas accretion from different sources in the Galactic potential and concludes that accretion from a diffuse halo is able to channel fresh ISM onto the most currently active star-forming Galactocentric radii, while accretion from dwarf satellites gas stripping would be dominant outside of the Solar circle therefore requiring efficient inward gas radial transport. While the latter is possible in principle, the need to transfer angular momentum outward would seem to make the process very inefficient even in the presence of well developed spiral arms (e.g., \citealt{Peek2009}). 

These flows could also correspond to the High-Velocity or Intermediate-Velocity Clouds observed in the HI 21-cm data \citep{Putman+2012}, with densities of the order of 0.1\,\cmthree\ and line-of-sight-velocities of hundreds of \kms, again compatible with the values required to cause and maintain local disk distortions (eq. \ref{flowdens}). As the disk distortions are mostly seen as bendings toward negative latitudes, the proposed mechanism would require that most of the incoming  gas comes from northern Galactic latitudes. Such an asymmetry is indeed observed in extraplanar gas in the  major IVCs \cite{Marasco+2012} that are believed to be a manifestation of the fountain gas, as well as in the major HVC that in part could be due to halo gas flows.

%  Such densities are lower than the diffuse ISM in the disk, but the velocities are only mildly in excess of the disk rotational speed.  However, flows of atomic gas from the intergalactic medium accreting onto the disk would not have the episodic behaviour of fountain gas, so that the interaction with the disk could last for a longer time than the one implied by the length-scale of the observed disk distortions. The observed distortions could then be interpreted as the result of a relatively long-duration and nearly stationary disk-flow interaction.

%The formation of dense and filamentary molecular clouds in the post-shock regions of large-scale flows has been among the most favorite mechanisms to explain the properties of filamentary star-forming regions both with active formation in intermediate and high-mass regimes (e.g. DR21, ***references***) as well as in relatively earlier IRDC stage (e.g. Peretto et al. 2013). Supporting evidence for this shocking flows is provided by large-scale detection of SiO vibrational lines around filamentary star-forming regions and IRDCs; motions related to spiral arms have been proposed as driving agents for these flows. The possibility that accreting vertical flows not only distort the gaseous component of the Galactic disk as observed by Herschel, the diffuse as well as the dense and star-forming, but indeed constitute the bulk of the flowing material that shocks the disk and forms the dense star-forming structures, will be investigated in forthcoming papers.

\subsection{Evolution of the distortions}

As soon as stars are formed out of the cold and dense clouds their cross-section is reduced, such that they can dynamically decouple from their gaseous environment and rapidly settle down onto the $b=0$\adeg\ midplane. The representative maximum displacement of the barycenter of the diffuse 500-\um\ emission as well as of the latitude medians of the 70- and 250-\um\ compact sources with respect to the $b=0$\adeg\ midplane is $\sim $ 0\adeg .2. At a distance of 5\,kpc, this is a representative value for the portion of Sagittarium Arm undergoing the downward bend in the 1$^{st}$ Quadrant (\S\ref{dist_1}), and corresponds to about 17\,pc. With a vertical velocity of 7\,km\,s$^{-1}$, comparable to the Sun's vertical speed with respect to the LSR (e.g., \citealt{bm98}), an object would cover this space in about 2.3 Myr. At the $\sim$8.4\,kpc distance that we assume on average for the Galactic Bar, where the spectacular 25\adeg$\geq l  \geq-25$\adeg\ sinusoidal-like latitude modulation is seen (\S\ref{dist_bar}b), the vertical displacement would be of the order of 30\,pc, requiring $\sim$4\,Myr to reach the midplane at the same 7-km\,s$^{-1}$ vertical speed. 

These timescales are compatible with the formation timescale of the low- and intermediate-mass stars that constitute the bulk of the stellar mass in the clusters that form in the \higal\ clumps. It is then plausible that the observed distortions in the cold and star-forming component of the Galactic disk can settle down toward the flatter latitude distribution seen for the mix of relatively more evolved YSOs and MS/post-MS objects traced at by MIPSGAL 24\um\ or WISE 22\um\ sources, and for the stellar Galactic disk in general. This would be the first direct evidence of a steady star formation from accretion of fresh extraplanar material.

%This qualitative argument shows that, in a time comparable to the formation timescale of the low- and intermediate-mass stars that constitute the bulk of the stellar mass in the clusters that form in the \higal\ clumps, such distortions in the cold and star-forming component of the Galactic disk can settle down toward the flatter latitude distribution seen for the mix of relatively more evolved YSOs and MS/post-MS objects traced at 24\,\um\ in MIPSGAL sources, and the stellar Galactic disk in general. This would be the first direct evidence (other than the obvious fact that the Milky Way has been forming stars for much longer than the past 10$^9$ years) of a steady accretion of fresh material from the Galactic halo.

\section{Conclusions}

The assembly of panoramic views of the inner Milky Way from the \textit{Herschel}\ \higal\ survey enables studies of the spatial distribution in Galactic longitude and latitude of the interstellar medium and of dense star-forming clumps with unprecedented detail and statistical significance. 

The width of the Galactic Plane expressed as the FWHM of the latitude distribution of high-reliability \higal\ compact sources, with counterpart in at least three adjacent {\it Herschel} photometric bands, has a mean value of 0\adeg\!.75 over the +60\adeg $\geq l \geq -60$\adeg\ inner Galaxy.  The width of the Plane measured by the FWHM of the diffuse 500-\um\ emission from dust is about  twice as large.  The peak of the overall latitude distribution of \higal\ sources is at $b\sim -$0\adeg\!.06, essentially coincident with the results from ATLASGAL. 

Large-scale latitude distortions of the Galactic Plane are visible in the $\sim$130\adeg -long 500-\um\ mosaic presented in fig. \ref{barpan}. A quantitative determination of the peak and width of the latitude distribution of the dust column density has been made via a polynomial fit to the emission in bins of 1\adeg\ in longitude.  Strong star-forming complexes are clipped out to obtain a more reliable determination of the distribution peak in latitude. The width of the distribution is determined by the half-power width of the fit.
%latitudes where the emission is 50\%\ of the peak value, essentially coinciding with the FWHM of the distribution as the emission brightness  in the peak position is a factor of a few brighter than at the extreme latitude locations in the $b$-strip imaged with  \higal. 
The number density of compact sources from the band-merged \higal\ photometric catalogues is analysed as a function of latitude, by computing the median latitude values for sources in 1\adeg-wide longitude bins. The detailed latitude distribution as a function of longitude shows clear modulations visible, both for the diffuse emission and for the compact sources, as large-scale bending modes over all of the considered longitude range. Bends are mostly toward negative latitudes, with excursions of $\sim$0\adeg\!.2 below the midplane in the smoothed distributions peaking at $l \sim$ +40\adeg, +12\adeg, $-$25\adeg\ and $-$40\adeg. The only positive bend peaks at $l\sim -$5\adeg. No comparable modulations can be found in the distributions of the entire samples of MIPSGAL 24-\um\ or the WISE 22\um\ point source distribution analysed with the same methodology, and none is reported for the GLIMPSE point-source distribution. The analogous distribution using the subsample of MIPSGAL/GLIMPSE sources selected based on color criteria targeted toward YSOs \citep{Robitaille2008}, exhibits instead some of the features shown by the Herschel sources.

The distortions of the Galactic inner disk revealed by {\it Herschel} confirm previous findings with much lower statistical significance from CO surveys and HII/OB source counts. The fact that no such distortions are visible with tracers of more evolved YSOs or the stellar disk in general would rule out gravitational instabilities or satellite-induced perturbations, as they should act on both the diffuse and stellar disk components. We propose that incoming flows of diffuse material from the Galactic halo interact with the disk causing the bends seen in the {\it Herschel} data. These effects are still visible in the distribution of very young star-forming clumps. Stars have a much lower cross-section with the supposed incoming flows, therefore decoupling from them and relaxing onto the stellar disk. The timescale required for the disappearance of the distortions from the diffuse ISM to the relatively evolved YSO stages are compatible with star formation timescales, assuming a velocity equal to the Sun's vertical speed in the disk.

\begin{acknowledgement}
We thank F. Fraternali and the referee, Dr. Mark Heyer, for their valuable comments that improved the paper. This work is part of the VIALACTEA Project, a Collaborative Project under Framework Programme 7 of the European Union, funded under Contract \# 607380 that is hereby acknowledged. We also thank ASI, Agenzia Spaziale Italiana, for its past support to the Hi-GAL project under contracts I/038/080/0 and I/029/12/0.

{\it Herschel} is an ESA space observatory with science instruments provided by European-led Principal Investigator consortia and with important participation from NASA. 

PACS has been developed by a consortium of institutes led by MPE (Germany) and including UVIE (Austria); KUL, CSL, IMEC (Belgium); CEA, OAMP (France); MPIA (Germany); IAPS, OAP/OAT, OAA/CAISMI, LENS, SISSA (Italy); IAC (Spain). This development has been supported by the funding agencies BMVIT (Austria), ESA-PRODEX (Belgium), CEA/CNES (France), DLR (Germany), ASI (Italy), and CICYT/MCYT (Spain). 

SPIRE has been developed by a consortium of institutes led by Cardiff Univ. (UK) and including Univ. Lethbridge (Canada); NAOC (China); CEA, LAM (France); IAPS, Univ. Padua (Italy); IAC (Spain); Stockholm Observatory (Sweden); Imperial College London, RAL, UCL-MSSL, UKATC, Univ. Sussex (UK); Caltech, JPL, NHSC, Univ. Colorado (USA). This development has been supported by national funding agencies: CSA (Canada); NAOC (China); CEA, CNES, CNRS (France); ASI (Italy); MCINN (Spain); Stockholm Observatory (Sweden); STFC (UK); and NASA (USA).

\end{acknowledgement}

%\bibliography{/Users/molinari/Science/sergio_bib}
%\bibliographystyle{/Users/molinari/Science/aa/aa}

\end{document}